\documentclass[10pt,prl,superscriptaddress,preprintnumbers,balancelastpage,twocolumn]{revtex4-2}

\usepackage{graphicx}
\usepackage{epstopdf}
\usepackage{bm}
\usepackage[utf8]{inputenc}
\usepackage[hidelinks,colorlinks]{hyperref}
\usepackage{color}
\usepackage{soul}
\usepackage{amsmath}
\usepackage{amssymb}
\usepackage{mathtools}
\usepackage{xfrac}
\usepackage{rotating}
\usepackage{enumitem}
\usepackage{scrextend}
\usepackage{slashed}
\usepackage{subfig}
\usepackage{braket}

\def\beq{\begin{equation}}
	\def\eeq{\end{equation}}
\def\beqa{\begin{eqnarray}}
	\def\eeqa{\end{eqnarray}}

\def\be{\begin{equation}}
	\def\ee{\end{equation}}
\def\bea{\begin{eqnarray}}
	\def\eea{\end{eqnarray}}

\def\nn{\nonumber}

\newcommand{\bei}{\begin{itemize}}
\newcommand{\eei}{\end{itemize}}

\newcommand{\T}{{\bf T}}

\newcommand{\tts}{\mathbf{T}_t^2}
\newcommand{\tsu}{\mathbf{T}_{s-u}^2}

\newcommand{\al}{\alpha}

\newcommand{\eps}{\epsilon}

\newcommand{\mm}{\mathcal{M}_{ij\to ij}}

\newcommand{\mExpM}[2]{\hat{\mathcal{M}}^{(-,#1,#2)}}

\renewcommand\arraystretch{1.5}

\allowdisplaybreaks

\begin{document}

\begin{flushright}
CERN-TH-2021-225 
\vspace*{-25pt}
\end{flushright}

\title{Disentangling the Regge cut and Regge pole in perturbative QCD}

\author{Giulio Falcioni}
\email{Giulio.Falcioni@ed.ac.uk}
\affiliation{Higgs Centre for Theoretical Physics, School of Physics and Astronomy, The University of Edinburgh, Edinburgh EH9 3FD, Scotland, UK}

\author{Einan Gardi}
\email{Einan.Gardi@ed.ac.uk}
\affiliation{Higgs Centre for Theoretical Physics, School of Physics and Astronomy, The University of Edinburgh, Edinburgh EH9 3FD, Scotland, UK}

\author{Niamh Maher}
\email{N.Maher@sms.ed.ac.uk}
\affiliation{Higgs Centre for Theoretical Physics, School of Physics and Astronomy, The University of Edinburgh, Edinburgh EH9 3FD, Scotland, UK}

\author{Calum Milloy}
\email{CalumWilliam.Milloy@unito.it}
\affiliation{Dipartimento di Fisica and Arnold-Regge Center, Universit\`{a} di Torino,
  and INFN, Sezione di Torino, Via P. Giuria 1, I-10125 Torino, Italy}

\author{Leonardo Vernazza}
\email{Leonardo.Vernazza@to.infn.it}
\affiliation{Dipartimento di Fisica and Arnold-Regge Center, Universit\`{a} di Torino,
and INFN, Sezione di Torino, Via P. Giuria 1, I-10125 Torino, Italy}
\affiliation{Theoretical Physics Department, CERN, Geneva 1211, Switzerland}
\date{\today}

\begin{abstract}
The high-energy limit of gauge-theory amplitudes features both a Regge pole and Regge cuts. We show how to disentangle these, and hence how to determine the Regge trajectory beyond two loops. While the nonplanar part of multiple Reggeon $t$-channel exchange forms a Regge cut, the planar part contributes to the pole along with the single Reggeon. With this, we find that the infrared singularities of the trajectory are given by the cusp anomalous dimension. By matching to recent QCD results, we determine the quark and gluon impact factors to two loops and the Regge trajectory to three loops.
\end{abstract}

\maketitle

The high-energy behaviour of QCD scattering amplitudes has long been a source of inspiration and unique insight into the gauge dynamics~\cite{Grisaru:1973wbb,Grisaru:1973vw,Grisaru:1973ku,Lipatov:1976zz}. Specifically, gluon Reggeization and the BFKL equation~\cite{Fadin:1975cb,Kuraev:1976ge,Kuraev:1977fs,Balitsky:1978ic} connect between fixed-order computations of partonic scattering and Regge theory~\cite{Gribov:1961fr,Amati:1962nv,Mandelstam:1963cw,Eden:1966dnq,Collins:1977jy,Gribov:1967vfb}, where the high-energy asymptotic behaviour of amplitudes is described in terms of its singularities in the complex angular momentum plane. 

Consider a $2\to 2$ scattering amplitude $\mm$ of massless partons, $i(p_1)+j(p_2)\to j(p_3)+i(p_4)$, described in terms of Mandelstam variables $s \equiv (p_1 + p_2)^2$, $t \equiv (p_1 - p_4)^2$ and $u \equiv (p_1 - p_3)^2 =-s-t$, in the high-energy limit, $s \gg -t$. At the leading-logarithmic (LL) approximation, rapidity logarithms, $\log ({s}/{(-t)})$, appearing in the one-loop amplitude simply exponentiate as $\left({s}/({-t})\right)^{C_A\alpha_g}$, forming a Regge pole. This can be seen as due to the $t$-channel exchange of a single \emph{Reggeized gluon}, or Single Reggeon (SR)~\cite{Fadin:2020lam}. The exponent is the leading-order gluon Regge trajectory, $\alpha_g(t)= \al_g^{(1)} a+{\cal O}(a^2)$ with $a\equiv\alpha_s(-t)/\pi$ and
in $d=4-2\epsilon$ dimensions,
\begin{align}
\label{al_g1}
\al_g^{(1)} = \frac{r_\Gamma}{2\eps}\,;
\qquad\quad 
r_\Gamma = e^{\eps\gamma_E}
\frac{\Gamma^2(1-\eps)\Gamma(1+\eps)}{\Gamma(1-2\eps)}\,.
\end{align}
The colour structure of the LL amplitude remains the same as at tree level, a pure octet exchange in the $t$ channel. Beyond LLs other colour states contribute where multiple Reggeon (MR) $t$-channel exchanges take place. 

At higher logarithmic accuracy, it is useful to split the amplitude into definite \hbox{$s\leftrightarrow u$} signature components:
\be
\mm = \,\,\mm^{(-)} + \mm^{(+)}.
\ee 
Further defining a signature-symmetric logarithm
\beq
\label{eq:siglog}
L\equiv \log\left(\frac{s}{-t}\right)-\frac{i\pi}{2}=\frac12\left[\log\left(\frac{-s}{-t}\right)+\log\left(\frac{-u}{-t}\right)\right],
\eeq
and expanding the amplitude as 
\begin{equation}\label{eq:expansionDef}
	\mm^{(\pm)} = \sum_{n=0}^\infty 
	a^n \sum_{m=0}^nL^m\mm^{(\pm,n,m)},
\end{equation}
one can show~\cite{Caron-Huot:2017fxr} that the coefficients of the odd (even) amplitude, $\mm^{(\mp,n,m)}$, are purely real (imaginary). Moreover, $\mm^{(-)}$ only receives contributions from $t$-channel exchange of an odd number of Reggeons, while $\mm^{(+)}$, an even number. Consequently, at the next-to-leading logarithmic (NLL) accuracy\, the odd amplitude is still governed by a SR~\cite{Fadin:2006bj,Ioffe:2010zz,Fadin:2015zea}, which is proportional to the tree amplitude $\mm^{\text{tree}}=g_s^2\, \frac{2s}{t} \, T_i \cdot T_j$ and factorizes as
\begin{equation}
\label{eq:SR}
\mm^{(-)\, \rm SR} = C_i(t)\,C_j(t)\,e^{\alpha_g(t)\,C_A\,L}\,\mm^{\text{tree}} \,,   
\end{equation}
where the \emph{impact factors} $C_{i/j} (t)= 1+ C_{i/j}^{(1)} a+\ldots$ associated with the parton $i/j$, need to be computed to one loop, while the gluon Regge trajectory must be computed to two loops, $\alpha_g(t)= \al_g^{(1)} a+\al_g^{(2)} a^2+ {\cal O}(a^3)$. 
From the Regge-theory perspective eq.~(\ref{eq:SR}) manifests a pure Regge pole.
The signature-even amplitude is instead governed by the exchange of a pair of Reggeised gluons, which forms a Regge cut~\cite{Kuraev:1977fs,Balitsky:1978ic,Caron-Huot:2013fea,Caron-Huot:2017zfo,Caron-Huot:2020grv}. 

A significant increase in complexity arises upon considering the odd amplitude at the next-to-next-to-leading logarithmic (NNLL) approximation, which features both a Regge pole and a Regge cut~\cite{DelDuca:2001gu,DelDuca:2013ara,DelDuca:2013dsa,DelDuca:2014cya,Caron-Huot:2017fxr,Fadin:2016wso,Fadin:2017nka,Fadin:2020lam,Fadin:2021csi,Falcioni:2020lvv,Falcioni:2021buo}. Since the high-energy analytic properties are only manifest upon resumming the entire perturbative series, it is not at all obvious how to disentangle the Regge pole from the Regge cut in an order-by-order computation. This paper presents a solution to this fundamental question.  

\begin{figure}[htb]
    \centering
    \subfloat[\label{fig:h33DL}]{\includegraphics[scale=1.1]{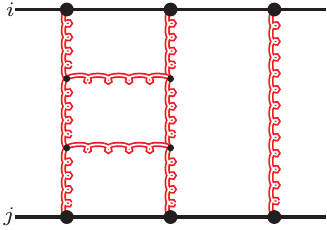}}
    \hspace{30pt}
    \subfloat[\label{fig:h33ML}]{\includegraphics[scale=1.1]{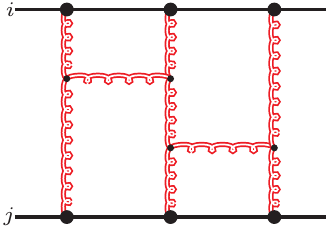}}
    \caption{\label{ladder_graphs}Four-loop ladder diagrams contributing to the three-Reggeon transition amplitude $A^{(4)}_{3\to 3}$ in eq.~(\ref{eq:NNLLtower}). }
\end{figure}

\begin{figure}[h]
    \subfloat[\label{fig:h13-4}]{\includegraphics[scale=.38]{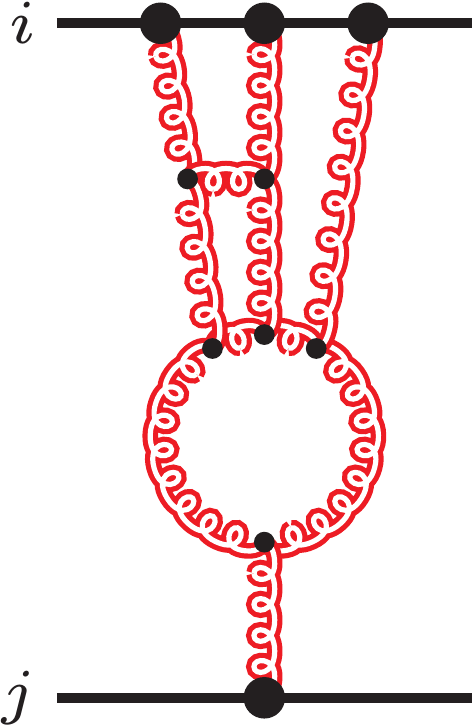}}\hspace{60pt}
    \subfloat[\label{fig:h3113a}]{\includegraphics[scale=.38]{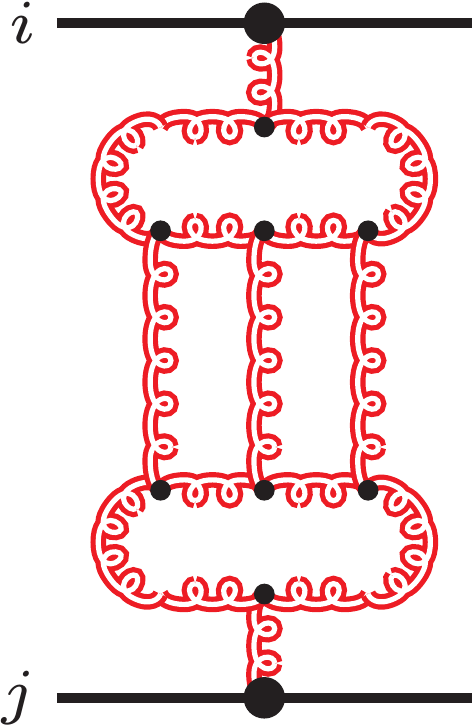}}
    \caption{Four-loop MR transitions involving triple and single Reggeon mixing, $A^{(4)}_{3\to 1}$ and $A^{(4)}_{1\to 3\to 1}$ in eq.~(\ref{eq:NNLLtower}).}
    \label{fig:mixing_graphs}
\end{figure}

Key to this progress is the possibility to directly compute the contributions to the scattering amplitude from the $t$-channel MR exchange. Specifically, the NNLL tower of MR corrections is governed by triple Reggeon exchange and its mixing with a single Reggeon~\cite{Caron-Huot:2017fxr,Fadin:2016wso,Fadin:2017nka,Fadin:2021csi,Falcioni:2020lvv,Falcioni:2021buo} and there is now an established method~\cite{Caron-Huot:2017fxr,Falcioni:2020lvv,Falcioni:2021buo}, based on the shockwave formalism~\cite{Caron-Huot:2013fea}, to evaluate these contributions through an iterative solution of the Balitsky-JIMWLK rapidity evolution equation~\cite{Balitsky:1995ub,Kovchegov:1999yj,JalilianMarian:1996xn,JalilianMarian:1997gr,Iancu:2001ad}.
The entire tower of NNLL corrections only requires the leading-order evolution kernel, thus yielding a universal result for all gauge theories~\cite{Falcioni:2020lvv,Falcioni:2021buo}. 
The result takes the form of generalised ladder diagrams (Figs.~\ref{ladder_graphs} and~\ref{fig:mixing_graphs}) giving rise to $2-2\epsilon$ dimensional integrals over transverse momenta with purely gluonic colour structures. 
Explicit computations in this framework have been recently performed through four loops, providing essential input for the present work.
  
The colour structure arising from MR exchange is rather complex. When considered for specific colour representations of the scattered partons, the $t$-channel colour flow of $\mm^{(-)\, \rm MR}$ involves the full set of representations compatible with the odd signature. Besides the octet, this includes the decuplet for gluon-gluon scattering and the singlet for quark-quark scattering. Rather than projecting the amplitude into colour representations, we use a basis of operators corresponding to colour flow through the three channels~\cite{Dokshitzer:2005ig,DelDuca:2011wkl,DelDuca:2011ae}:
\begin{align}
	\label{TtTsTu}
	\begin{split}
		\T_s &= \T_1+\T_2,\hspace{0.4cm}
		\T_u = \T_1+\T_3,\hspace{0.4cm}
		\T_t  = \T_1+\T_4,
	\end{split}
\end{align}
where ${\bf T}_{k}$ corresponds to the colour generator associated with the parton $k$. The MR contribution can then be expressed~\cite{Falcioni:2020lvv,Falcioni:2021buo} in terms of commutators of the operators $\tts$ and $\tsu\equiv \tfrac12 \left(\T_s^2-\T_u^2\right)$ along with quadratic and quartic Casimirs in the representation of the scattered partons and in the adjoint representation. This basis allows for a general treatment of high-energy $2\to 2$ scattering in any representation through NNLL, making signature symmetry manifest. It also provides a transparent separation between planar and nonplanar colour factors, as commutators of $\tts$ and $\tsu$ are nonplanar~\cite{Falcioni:2021buo}. 

The tower of NNLL MR contributions to the amplitude is most naturally expressed in terms of the \emph{reduced} amplitude~\cite{Caron-Huot:2017fxr,Falcioni:2020lvv,Falcioni:2021buo},  via 
\begin{equation}
\label{eq:MR}
\mm^{(-)\, \text{MR}} = Z_i\,Z_j\,e^{\alpha_g(t)\tts L}\hat{\mathcal{M}}_{ij\to ij}^{(-)\, \text{MR}},
\end{equation}
where 
\begin{equation}
	Z_i(t)=\exp\left\{-\frac{1}{2}\int_0^{\mu^2}\frac{d\lambda^2}{\lambda^2}\Gamma_i\left(\al_s(\lambda^2),\frac{-t}{\lambda^2}\right)\right\}
	\label{Zi}
\end{equation}
with the anomalous dimension 
\begin{align}\nonumber
	\Gamma_i\bigg(\alpha_s(\lambda^2),\frac{-t}{\lambda^2}\bigg)&=
	\Gamma^{\rm{cusp}}_i(\alpha_s(\lambda^2))\log \frac{-t}{\lambda^2}+2\gamma_i(\alpha_s(\lambda^2)),
\end{align} 
where $\Gamma^{\rm{cusp}}_i$ and $\gamma_i$ are, respectively, the 
cusp~\cite{Korchemsky:1987wg,Boels:2017ftb,Boels:2017skl,Moch:2017uml,Grozin:2017css,Henn:2019swt,vonManteuffel:2020vjv,Agarwal:2021zft} and the  collinear anomalous dimensions of parton $i$ \cite{Moch:2005tm,Gehrmann:2010ue,Dixon:2017nat,Falcioni:2019nxk,vonManteuffel:2020vjv,Agarwal:2021zft}. The reduced NNLL amplitude  takes the form~\cite{Falcioni:2020lvv,Falcioni:2021buo}:
\begin{align}
\label{eq:NNLLtower}
    &\hat{\mathcal{M}}_{ij\to ij}^{(-)\, \text{MR}}=\pi^2\sum_{n\geq 2}
    \left(ar_\Gamma\right)^n \,{(-L)^{n-2}}\bigg[\,{A}^{(n)}_{3\to3} \\
    &+ \Theta(n\geq 3)\,
    \left({A}^{(n)}_{1\to 3}+{A}^{(n)}_{3\to 1}\right)+ \Theta(n\geq 4) \nonumber {A}^{(n)}_{1\to3\to1}\bigg]\,,
\end{align}
where $\Theta$ is a heaviside function, and ${A}^{(n)}_{m\to k}$ corresponds to an $n$-loop transition amplitude describing the emission of $m$ Reggeons from the projectile $i$ followed by absorption of $k$ Reggeons by the target $j$. The tower of NNLL corrections to the odd amplitude is the first instance where transitions between states consisting of a different number of Reggeons occur~\cite{Caron-Huot:2017fxr} and refs.~\cite{Falcioni:2020lvv,Falcioni:2021buo} gained insight into their characteristics. All transition amplitudes in eq.~(\ref{eq:NNLLtower}) are universal for all ${\rm SU}(N_c)$ gauge theories and at $n$ loops they feature uniform transcendental weight~$n$ (considering $1/\epsilon$ as having weight $1$).
A~salient feature is the cancellation of planar corrections. Partial cancellation of these starts at three loops, where the terms ${A}^{(n)}_{3\to 1}$ and ${A}^{(n)}_{1\to 3}$, which are proportional to the quartic Casimirs in the $i$ and $j$ representations, respectively, completely cancels against similar contributions in ${A}^{(n)}_{3\to 3}$. An additional cancellation kicks in at four loops, where the $A^{(n)}_{1\to 3\to 1}$ transition first appears (Fig.~\ref{fig:mixing_graphs} (b)), upon which no planar contributions survive. Explicitly shown at four loops~\cite{Falcioni:2021buo}, this complete cancellation conjecturally extends to all higher orders,~$n\geq 4$: planar corrections are only present in $\hat{\mathcal{M}}_{ij\to ij}^{(-)\, \text{MR}}$ at two and three loops, before all transition-amplitude channels open up.

A crucial observation is that while the Regge cut arises exclusively due to MR contributions to the amplitude, MR exchanges do contribute also to the Regge pole. 
The latter statement becomes evident in the large-$N_c$ limit, where it is known that the amplitude only features a Regge pole~\cite{Eden:1966dnq,Collins:1977jy}, and yet, MR contributions are present~\cite{Bartels:1980pe,Kwiecinski:1980wb,Lipatov:1993yb,Faddeev:1994zg,Derkachov:2001yn,Derkachov:2002wz,Caron-Huot:2017fxr}. 
It is also known that Regge cuts only arise due to nonplanar diagrams~\cite{Mandelstam:1963cw,Collins:1977jy}. This strongly suggests that the Regge cut should be identified as the nonplanar part of the MR contribution, while the Regge pole corresponds to SR plus the planar MR contributions. We will show that the properties of the computed MR contributions precisely match this conclusion. Indeed, while the nonplanar part of the MR contribution involves different colour-flow components, depending on the nature of the scattered partons, its planar part, present at two and three loops, is restricted to the octet $t$-channel exchange, it is entirely independent of the scattered partons, and thus naturally forms part of the Regge pole, along with the SR contribution:
\begin{align}
\label{pole-cut}
\mm^{(-)} &\!= \underbrace{\mm^{(-)\,{\rm SR}} +
\left.\mm^{(-)\,{\rm MR}}\right\vert_{\text{planar}}} 
\!\!+\!
\left.\mm^{(-)\,{\rm MR}}\right\vert_{\text{nonplanar}} \nn \\
&\!= \hspace*{40pt} \mm^{(-)\,{\rm pole}} \hspace*{30pt} + \hspace*{1pt} \mm^{(-)\,{\rm cut}}.
\end{align}
This separation of the amplitude will be referred to as the cut scheme. At NNLL accuracy, the pole contribution is uniquely fixed by the impact factors through two loops and the gluon Regge trajectory at three loops. Thus, the inherent nonplanar nature of the NNLL MR contribution at four loops (and above) is essential.

The nonplanar part of the NNLL MR contribution to the amplitude, which constitutes the Regge cut in eq.~(\ref{pole-cut}), has the following expansion coefficients~\cite{Falcioni:2021buo}:
\begin{align}
\label{M20cut}
    &{\cal M}^{(-,2,0)\,\text{cut}}_{ij\to ij} = \pi^2(r_\Gamma)^2S^{(2)}(\eps)\left[(\tsu)^2-\frac{C_A^2}{4}\right] {\cal M}_{ij\to ij}^{\text{tree}}\,,\nn
    \\
  & {\text{with}}\quad  S^{(2)}(\epsilon)  = -\frac{1}{8\epsilon^2}+\frac{3}{4}\epsilon\zeta_3+\frac{9}{8}\epsilon^2\zeta_4+{\cal{O}}(\epsilon^3),
\end{align}
\begin{align}
\label{M31cut}
    &{\cal M}^{(-,3,1)\,\text{cut}}_{ij\to ij} =\, -\pi^2(r_\Gamma)^3\bigg[S_A^{(3)}(\eps)\tsu[\tsu,\tts]
    \\& +S_B^{(3)}(\eps)[\tsu,\tts]\tsu\bigg]{\cal M}_{ij\to ij}^{\text{tree}}
    +\al_g^{(1)}\tts{\cal M}^{(-,2,0)\,\text{cut}}_{ij\to ij},\nn
\end{align}
with 
\begin{subequations}
\label{SAB}
\begin{align}
    S^{(3)}_A(\epsilon) &= \frac{1}{48\epsilon^3} + \frac{37}{24}\zeta_3 + \frac{37}{16}\epsilon\,\zeta_4 + {\cal{O}}(\epsilon^2),
\\
    S^{(3)}_B(\epsilon) &= \frac{1}{24\epsilon^3} + \frac{1}{12}\zeta_3 + \frac{1}{8}\epsilon\,\zeta_4 + {\cal{O}}(\epsilon^2),
    \end{align}
\end{subequations}
and for $n\geq 4$
\begin{align}\label{eq:McutAllOrder}
    {\cal M}^{(-,n,n-2)\,\text{cut}}_{ij\to ij}&=\sum_{m=0}^{n-4}\frac{1}{m!}\left(\al_g^{(1)}\tts\right)^m\mExpM{n-m}{n-m-2}\nn\\&\hskip-14pt+\frac{1}{(n-3)!}\left(\al_g^{(1)}\tts\right)^{n-3}{\cal M}^{(-,3,1)\,\text{cut}}_{ij\to ij}
    \\& \hskip-14pt-\frac{n-3}{(n-2)!}\left(\al_g^{(1)}\tts\right)^{n-2}{\cal M}^{(-,2,0)\,\text{cut}}_{ij\to ij}.\nn
\end{align}

The planar part of the NNLL MR contribution in eq.~(\ref{pole-cut}) is
\begin{eqnarray}
\label{eq:MRplanar}
\begin{split}
&\hspace{-25pt} \left.\mm^{(-)\,\text{MR}}\right|_\text{planar} = \frac{\pi^2 a^2r_\Gamma^2 N_c^2}{6}  \,\mm^\text{tree}\bigg\{{S^{(2)}(\epsilon)}&
    \\   \hspace*{25pt}
&\hspace{-15pt}   -a r_\Gamma N_c L  \left[ \frac{1}{3}\left(S^{(3)}_A(\epsilon)-S^{(3)}_B(\epsilon)\right)-\frac{1}{2\epsilon}{S^{(2)}(\epsilon)} \right]\bigg\}, &
   \end{split}
\end{eqnarray}
with $S^{(2)}$ given in eq.~(\ref{M20cut}) and $S_A^{(3)}$ and $S_B^{(3)}$ in eq.~(\ref{SAB}). Terms of ${\cal{O}}(\alpha_s^4)$ in eq.~(\ref{eq:MRplanar}) vanish identically. Importantly, the result in eq.~(\ref{eq:MRplanar}) is proportional to the tree amplitude and is entirely independent of the scattered partons. Hence, using it in eq.~(\ref{pole-cut}) along with the SR contribution of eq.~(\ref{eq:SR}), the complete Regge-pole contribution factorizes as
\begin{equation}
\label{eq:tilde}
\mm^{(-)\, {\rm pole}} = \tilde{C}_i(t)\,\tilde{C}_j(t)\,e^{\tilde{\alpha}_g(t)\,C_A\,L}\,\mm^{\text{tree}} \,,   
\end{equation}
and one can deduce the relation between the two-loop impact-factor coefficients in eq.~(\ref{eq:tilde}) and those in eq.~(\ref{eq:SR}), 
\begin{align}
 \begin{split}
    \tilde{C}_{i/j}^{(2)} 
    \, = &\, C_{i/j}^{(2)} + N_c^2(r_\Gamma)^2\,\frac{\pi^2}{12} S^{(2)}(\epsilon),
    \end{split}
    \label{eq:Ctilde}
\end{align}
and similarly for the three-loop Regge trajectory, 
\begin{equation}
    \label{eq:reggeTilde}
    \tilde{\al}_g^{(3)} =\, \al_g^{(3)} - (r_\Gamma)^3N_c^2\frac{\pi^2}{18}\left(S^{(3)}_A(\epsilon)-S^{(3)}_B(\epsilon)\right).
\end{equation}
The coefficients $C_{i/j}^{(2)}$ and $\al_g^{(3)}$ were determined in ref.~\cite{Caron-Huot:2017fxr} from fixed-order calculations that had been available then. Here we shall use the cut scheme instead, and match it against state-of-the-art computations in QCD. 

At this point we have fixed the definition of all parameters entering the 
Regge-pole term in eq.~(\ref{pole-cut}) through NNLL. Since we expect this pole-cut separation to capture all-order analytic properties, it is interesting to examine the ensuing infrared singularity structure of the separate pole and cut terms. 
To this end let us define
\begin{equation}
\label{eq:Kcuspdef}
C_A\hat{\tilde{\alpha}}_g (t)\equiv 
C_A\tilde{\alpha}_g (t)+\frac{1}{2}\int_0^{\mu^2}\frac{d\lambda^2}{\lambda^2}
\Gamma^{\rm{cusp}}_A(\alpha_s(\lambda^2)) \,. 
\end{equation}
This definition is motivated by the insightful observation~\cite{Korchemskaya:1994qp, Korchemskaya:1996je} that the singularities of the gluon Regge trajectory are prescribed by the cusp anomalous dimension in the adjoint representation, such that $C_A\hat{\tilde{\alpha}}_g (t)$ is \emph{finite}. So far this has only been known to hold through two loops. We shall verify that it holds in the cut scheme at three loops. It remains a conjecture beyond this order. 

One can write the Regge-pole term in terms of an $s\leftrightarrow u$ symmetric sum of exponentials~\cite{Fadin:1993wh} rather than a single exponential of a symmetric logarithm. When this is done, one may define a collinear-subtracted impact factor $\bar{D}_i(t)$, which is \hbox{\emph{finite}}~\cite{Falcioni:2021buo,DelDuca:2014cya}:  
\beq
\label{eq:impactIRfact}
\tilde{C}_{i}(t) 
\equiv Z_{i}(t)\bar{D}_{i}(t)\sqrt{\cos\left(\frac{\pi C_A\tilde{\alpha}_g(t)}{2}\right)},
\eeq
where $Z_i(t)$ is given in eq.~(\ref{Zi}). We can then write an elegant expression for the odd amplitude at NNLL accuracy as follows:
\begin{widetext}
\begin{align}
\label{eq:colsubpole-vit-scheme}
   & \mathcal{M}^{(-)}_{ij\to ij} = Z_i(t)\,\bar D_i(t)\,Z_j(t)\,\bar D_j(t)\left[\left(\frac{-s}{-t}\right)^{C_A\tilde\alpha_g(t)} + \left(\frac{-u}{-t}\right)^{C_A\tilde\alpha_g(t)}\right]\mathcal{M}^{\text{tree}}_{ij\to ij}
    +\sum_{n=2}^\infty 
	a^n L^{n-2} \mm^{(-,n,n-2)\,{\rm cut}},
\end{align}
\end{widetext}
where the second term, representing the Regge cut, is nonplanar, and its 
coefficients $\mm^{(-,n,n-2)\, {\rm cut}}$ are given in eqs.~(\ref{M20cut},\ref{M31cut}) for $n=2,3$, and in eq.~(\ref{eq:McutAllOrder}) for $n\geq 4$. The four-loop reduced amplitude $\mExpM{4}{2}$ entering eq.~(\ref{eq:McutAllOrder}) is given in eq. (5.32) in ref.~\cite{Falcioni:2021buo}, while higher orders have not yet been computed.

Eq.~(\ref{eq:colsubpole-vit-scheme}) displays a transparent infrared-singularity structure: in the pole term such singularities are fully captured by the $Z_{i/j}$ factors and the divergences of the trajectory given by the integral over the cusp anomalous dimension in eq.~(\ref{eq:Kcuspdef}). Additional singularities are present in the cut term, but we note (see eqs.~(\ref{M20cut}) and~(\ref{M31cut})) that these do not feature any single $1/\epsilon$ pole at two and three loops, in line with the properties of the soft anomalous dimension~\cite{Caron-Huot:2017fxr,Almelid:2015jia,Almelid:2017qju,Falcioni:2020lvv,Falcioni:2021buo}. 
A $1/\epsilon$ pole does occur at ${\cal O}(\alpha_s^4L^2)$, which constitutes 
the first contribution to the soft anomalous dimension from the NNLL tower~\cite{Falcioni:2020lvv,Falcioni:2021buo}.

Having established the separation between the Regge cut and pole terms, we now compare our results to state-of-the-art two-~\cite{Ahmed:2019qtg} and three-loop~\cite{Caola:2021rqz} computations in QCD to extract the two-loop 
impact-factor coefficients for quarks and gluons to ${\cal O}(\epsilon^2)$ and the two- and three-loop Regge trajectory coefficients to ${\cal O}(\epsilon^2)$ and ${\cal O}(\epsilon^0)$, respectively. 
We performed the colour algebra with~\cite{Sjodahl:2012nk} and the required analytic continuations using ref.~\cite{Anastasiou:2013srw,Duhr:2019tlz}.
In line with eq.~(\ref{al_g1}), we get
$
\hat{\tilde{\al}}_g^{(1)}=\frac{1}{2\epsilon}({r_\Gamma-1})
$. 

We extract the two-loop Regge trajectory to ${\cal{O}}(\epsilon^2)$ by expanding eq.~(\ref{eq:colsubpole-vit-scheme}) for $gg\to gg$ to two loops and comparing it to the high-energy limit of the corresponding QCD amplitude computed in ref.~\cite{Ahmed:2019qtg}, obtaining
\begin{widetext}
\begin{align}
\label{eq:hattildea2}
    \hat{\tilde{\al}}_g^{(2)} &= C_A \left(\frac{101}{108}-\frac{\zeta_3}{8}\right)-\frac{7 n_f}{54} +\eps \bigg[C_A \Big(\frac{607}{324}-\frac{67 \zeta_2}{144}
    -\frac{33 \zeta_3}{8}-\frac{3 \zeta_4}{16}\Big)+n_f \left(-\frac{41}{162}+\frac{5 \zeta_2}{72}+\frac{3 \zeta_3}{4}\right)\bigg]
   \\&
    +\eps^2 \bigg[C_A \bigg(\frac{911}{243}-\frac{101 \zeta_2}{108}-\frac{1139 \zeta_3}{108}-\frac{2321 \zeta_4}{384}+\frac{41 \zeta_5}{8}
    +\frac{71 \zeta_2 \zeta_3}{24}\bigg) +n_f \left(\frac{7 \zeta_2}{54}+\frac{85 \zeta_3}{54}+\frac{211 \zeta_4}{192}-\frac{122}{243}\right)\bigg] \nn
    + \mathcal{O}(\eps^3).
\end{align}
\end{widetext}
Here, the terms of ${\cal{O}}(\epsilon^0)$ agree with the literature \cite{Fadin:1995km,Fadin:1995xg,Fadin:1996tb,Blumlein:1998ib,DelDuca:2001gu} and higher orders in $\epsilon$ are new results. Upon taking the planar limit, these agree with the recent calculation in ref.~\cite{DelDuca:2021vjq}.
We also verified that eq.~(\ref{eq:hattildea2}) agrees with the gluon Regge trajectory extracted by comparing the expansion of eq.~(\ref{eq:colsubpole-vit-scheme}) for quark-quark scattering with the high-energy limit of the corresponding amplitude in ref.~\cite{Ahmed:2019qtg}. Furthermore, from the same comparison we determine the quark and gluon impact factors, $\tilde{C}_{q}(t)$ and $\tilde{C}_{g}(t)$ to two loops through $\mathcal{O}(\eps^2)$. These are conveniently expressed in terms of the corresponding collinear-subtracted impact
factors $\bar{D}_i(t)$ defined in eq.~(\ref{eq:impactIRfact}), which we expand as $\bar{D}_i(t) = 1+a \bar{D}_i^{(1)} + a^2 \bar{D}_i^{(2)}+ \dots$. 
The one-loop impact factors are 
\begin{align}
\label{Dbargq1}
 \bar{D}_g^{(1)}&=C_A\left(\zeta_2-\frac{67}{72}\right)+\frac{5}{36}n_f+{\cal O}(\epsilon)\,,
 \\
 \bar{D}_q^{(1)}& =
C_A \bigg(\frac{85}{72} + \frac{3 \zeta_2}{4}\bigg)
+ C_F \bigg(\frac{\zeta_2}{4}-2\bigg)
-\frac{5}{36} n_f +{\cal O}(\epsilon)\,, \nn
\end{align}
while the two-loop ones are
\begin{widetext}
\begin{align}
\label{two_loop_impact_factors}
\bar{D}_g^{(2)}&= C_A^2\Big(\frac{335}{288}\zeta_2+\frac{11}{18}\zeta_3-\frac{3}{32}\zeta_4-\frac{26675}{10368}\Big)
    +C_A n_f\Big(\frac{49}{108}-\frac{25}{144}\zeta_2+\frac{5}{36}\zeta_3\Big)+C_F n_f\!\Big(\frac{55}{192}-\frac{\zeta_3}{4}\Big)
    -\frac{25}{2592}n_f^2+ {\cal O}(\epsilon)\,,
    \\
\bar{D}_q^{(2)}&= \nn
 C_A^2 \bigg(\frac{73 \zeta_2}{32} 
- \frac{43 \zeta_3}{48} 
- \frac{53 \zeta_4}{64} 
+\frac{13195}{3456} \bigg)
+C_A C_F \bigg(  
- \frac{5 \zeta_2}{2} 
+ \frac{475 \zeta_3}{144} 
+ \frac{65 \zeta_4}{32} 
- \frac{78229}{10368} \bigg) 
+ C_F^2 \bigg(
\frac{21 \zeta_2}{16} 
- \frac{15 \zeta_3}{8} \\ \nn
&\hspace{1.5cm} 
- \frac{83 \zeta_4}{64} 
+\frac{255}{128} \bigg) 
+ C_A n_f \bigg(
- \frac{5\zeta_2}{16} 
- \frac{7 \zeta_3}{24}
-\frac{385}{432} \bigg) 
+ C_F n_f \bigg(
 \frac{\zeta_2}{8} 
+\frac{19 \zeta_3}{72}
+\frac{505}{648} \bigg) 
+ \frac{25}{864}n_f^2  + {\cal O}(\epsilon)\,.
\end{align}
\end{widetext}
Here we truncated the expansion at ${\cal O}(\epsilon)$. Results for $\bar{D}_i^{(1)}$ 
through ${\cal O}(\epsilon^4)$ and $\bar{D}_i^{(2)}$ 
through ${\cal O}(\epsilon^2)$ are provided in the ancillary file, along with $\tilde{C}_i^{(1)}$ to all orders in $\eps$ and~$\tilde{C}_i^{(2)}$ through to ${\cal O}(\epsilon^2)$. The results for~$\tilde{C}_i^{(2)}$ beyond ${\cal O}(\epsilon^0)$ are new. Upon taking the planar limit of~$\tilde{C}_g^{(2)}$ we find agreement with the recent calculation in ref.~\cite{DelDuca:2021vjq}. One can further check that in the supersymmetric limit, where $C_F = n_f=C_A$, the results for $\tilde{C}_q^{(n)}$ and $\tilde{C}_g^{(n)}$, coincide for the terms of highest and next-to-highest
transcendental weight (weights $2n$ and $2n-1$, respectively), which is a consequence of a supersymmetry Ward identity~\cite{Kunszt:1993sd,Bern:2002tk,Bern:2003ck,DeFreitas:2004kmi,Bern:2002zk}. Furthermore, upon extracting the terms of highest weight, one recovers the gluon impact factor in planar~${\cal N}=4$ SYM~\cite{DelDuca:2008pj,DelDuca:2008jg}. This result also coincides with the~${\cal N}=4$ SYM gluon impact factor in full colour~\cite{Henn:2016jdu,Caron-Huot:2017fxr}, upon using the cut scheme.

Finally, we determine the three-loop gluon Regge trajectory, by comparing the expansion of eq.~(\ref{eq:colsubpole-vit-scheme}) for quark-quark scattering with the recent calculation of this amplitude in QCD~\cite{Caola:2021rqz}. We find
\begin{widetext}
\begin{align}
\label{alpha_g_3}
\hat{\tilde\al}_g^{(3)} =
&\, C_A^2 \bigg(\frac{297029}{93312}-\frac{799 \zeta_2}{1296}-\frac{833 \zeta_3}{216}-\frac{77 \zeta_4}{192}+\frac{5}{24} \zeta_2 \zeta_3+\frac{\zeta_5}{4}\bigg) +C_A n_f \bigg(\frac{103 \zeta_2}{1296}+\frac{139 \zeta_3}{144}-\frac{5 \zeta_4}{96}-\frac{31313}{46656}\bigg) \nn \\ 
&\, +C_F n_f \bigg(\frac{19 \zeta_3}{72}+\frac{\zeta_4}{8}-\frac{1711}{3456}\bigg) +n_f^2 \left(\frac{29}{1458}-\frac{2\zeta_3}{27}\right) + \mathcal{O}(\eps),
\end{align}
\end{widetext}
where $\mathcal{O}(\eps)$ corrections are yet unknown. Notably, for $n_f=0$ there are no ${1}/{N_c^2}$ corrections, in line with the expected maximally non-Abelian nature of this quantity. Eq.~(\ref{alpha_g_3}) agrees with the recent calculation in the planar limit~\cite{DelDuca:2021vjq}.

{\bf Conclusions.} We have shown how to disentangle Regge pole and cut contributions, making direct use of the non-planar origin of the cut. A key step is to recognise that the planar part of MR exchanges contributes to the Regge pole, leading to the separation of the amplitude according to eq.~(\ref{pole-cut}).

Essential to this progress is the availability of a method to directly compute the MR contribution~\cite{Caron-Huot:2017fxr,Falcioni:2020lvv,Falcioni:2021buo}, in which explicit calculations were recently pushed through to four loops. This recent step has proven vital in uncovering the inherent non-planar nature of the NNLL reduced amplitude beyond three loops. Having planar MR contributions at two and three loops \emph{precisely matches} the parameters available in factorising the Regge-pole contribution, eq.~(\ref{eq:tilde}), namely fixing the two-loop impact factors and the three-loop Regge trajectory. If higher-order planar MR corrections were to appear at four loops or beyond, this would have been in direct conflict with the non-planar nature of the Regge cut~\cite{Mandelstam:1963cw,Collins:1977jy}.

A natural question arises regarding the uniqueness of the separation in eq.~(\ref{pole-cut}) in as far as non-planar corrections are concerned. A distinct possibility exists at two loops to absorb an additional ${\cal O}(1/N_c^2)$ contribution, which \emph{depends} on the scattered partons, into the pole-term impact factors, as previously proposed in ref.~\cite{Fadin:2017nka} (see (5.47) in~\cite{Falcioni:2021buo}). Importantly, this has no impact on the Regge trajectory. Moreover, the trajectory is expected to be maximally non-Abelian, and hence it is uniquely fixed. Remarkably, we find that the singularities of the trajectory are given by the integral over the cusp anomalous dimension, consistently with the proposition of refs.~\cite{Korchemskaya:1994qp,Korchemskaya:1996je}. 

Our procedure to separate the Regge pole and cut contributions was immediately put to use with the recent availability of complete three-loop calculations~\cite{Caola:2021rqz}. With two-loop results to high order in $\epsilon$ at hand for both quark and gluon scattering~\cite{Ahmed:2019qtg}, and three-loop results available for quark scattering~\cite{Caola:2021rqz}, we were able to fix the NNLL Regge-pole parameters, providing a robust check of present and future amplitude calculations. 
Further insight is obtained upon comparing different gauge theories: the NNLL Regge-cut contribution is entirely universal, while the Regge-pole parameters depend on the underlying gauge theory and on the scattered partons. Upon taking the supersymmetric limit of the pole parameters and extracting terms of maximal weight, one recovers the known ${\cal N}=4$ SYM results~\cite{Henn:2016jdu,Caron-Huot:2017fxr,DelDuca:2008pj,DelDuca:2008jg}. 
The QCD impact factor and trajectory extracted here are of direct relevance to the extension of the BFKL framework to NNLL accuracy, as well as to the study of multi-leg amplitudes in a variety of kinematic limits.

{\bf Acknowledgements.}
We thank Fabrizio Caola, Amlan Chakraborty, Giulio Gambuti, Andreas von Manteuffel and Lorenzo Tancredi for communications regarding their work \cite{Caola:2021rqz} and for comparing with their unpublished results. We thank Gregory Korchemsky, Simon Caron-Huot and Vittorio Del Duca for communication and encouragement. 
EG and GF are supported by the STFC Consolidated Grant ‘Particle Physics at the Higgs Centre’.
GF is supported by the ERC Starting Grant 715049 ‘QCDforfuture’ with Principal Investigator Jennifer Smillie.
CM's work is supported by the Italian Ministry of University and Research (MIUR),
grant PRIN 20172LNEEZ.  LV is supported 
by Fellini Fellowship for Innovation at INFN, funded by the 
European Union's Horizon 2020 research programme under the 
Marie Sk\l{}odowska-Curie Cofund Action, grant agreement no. 754496.

\bibliography{ReggeRefs}
\bibliographystyle{apsrev4-1}

\end{document}